\documentclass[10pt,conference]{IEEEtran}
\IEEEoverridecommandlockouts
\usepackage{cite}
\usepackage{amsmath,amssymb,amsfonts}
\usepackage{algorithmic}
\usepackage{graphicx}
\usepackage[dvipsnames]{xcolor}
\usepackage{textcomp}
\usepackage{pythonhighlight}
\usepackage{balance}
\usepackage{inconsolata}
\usepackage[braket, qm]{qcircuit}
\def\BibTeX{{\rm B\kern-.05em{\sc i\kern-.025em b}\kern-.08em
T\kern-.1667em\lower.7ex\hbox{E}\kern-.125emX}}

\usepackage{todonotes}
\usepackage{booktabs}
\usepackage{listings}
\usepackage[ngerman,english]{babel}
\usepackage[shortcuts,acronym]{glossaries}
\usepackage{hyperref}

\definecolor{darkgray}{RGB}{100,100,100}
\lstdefinelanguage{cypher}{
	morekeywords=[1]{RETURN, MATCH, AS, COUNT, WHERE, NOT, EXISTS, ORDER, BY, LIMIT, UNION, DESC},
  keywordstyle=[1]{\color{darkgray}},
  emph=[2]{QuantumGate, QuantumGateX, QuantumGateY, QuantumGateZ, 
  QuantumPauliGate, QuantumGateH, ClassicIf},
  emphstyle=[2]{\color{Tan}},
  emph=[3]{QuantumBitReference, ClassicBitReference},
  emphstyle=[3]{\color{Green}},
  emph=[4]{QuantumMeasure},
  emphstyle=[4]{\color{Blue}},
  emph=[5]{QuantumBit, ClassicBit},
  emphstyle=[5]{\color{Orange}},
  emph=[6]{DFG},
  emphstyle=[6]{\color{Red}},
  emph=[7]{EOG},
  emphstyle=[7]{\color{Blue}},
  	sensitive=false,
	morecomment=[l]{//},
	morecomment=[s]{/*}{*/},
	morestring=[b]",
	stringstyle={\color{Fuchsia}},
}
\usetikzlibrary{calc}
\tikzset{
  part/.style={
    rectangle,
    rounded corners,
    text centered,
    minimum height=0.75cm,
    minimum width=2.4cm,
    draw=black, very thick
  }
}

\usepackage{pifont}
\newcommand{\yes}{\checkmark}
\newcommand{\no}{\ding{55}}

\newcommand{\cpg}[0]{\textit{cpg}}

\newcommand{\allClassical}{classical parts}
\newcommand{\classical}{classical part}
\newcommand{\quantumService}{quantum service}
\newcommand{\local}{local part}
\newcommand{\remote}{remote part}
\newcommand{\quantum}{quantum part}
\newcommand{\whole}{classical and quantum parts}

\newacronym{cpg}{CPG}{Code Property Graph}
\newacronym{qcpg}{QCPG}{Quantum Code Property Graph}
\newacronym{ast}{AST}{Abstract Syntax Tree}
\newacronym{cfg}{CFG}{Control-flow Graph}
\newacronym{eog}{EOG}{Evaluation Order Graph}
\newacronym{dfg}{DFG}{Data-flow Graph}
\newacronym{pdg}{PDG}{Program-dependence Graph}

\newcommand\qiskit{Qiskit}
\newcommand\qasm{OpenQASM}

\usepackage{ctable}

\begin{document}

\title{A Uniform Representation of Classical and Quantum Source Code for Static
  Code Analysis\\
  \thanks{\textcopyright 2023
IEEE. Personal use of this material is permitted. Permission from IEEE must be obtained for all other uses, in any
current or future media, including reprinting/republishing this material for advertising or promotional purposes,
creating new collective works, for resale or redistribution to servers or lists, or reuse of any copyrighted
component of this work in other works. 

    This research is supported by the Bavarian Ministry of Economic Affairs, 
  Regional Development and Energy with funds from the Hightech Agenda Bayern.
  
DOI: \href{https://doi.org/10.1109/QCE57702.2023.00115}{10.1109/QCE57702.2023.00115}}
}

\author{\IEEEauthorblockN{1\textsuperscript{st} Maximilian Kaul}
  \IEEEauthorblockA{\textit{Fraunhofer AISEC}\\
    Garching, Germany\\
    \href{mailto:maximilian.kaul@aisec.fraunhofer.de}{maximilian.kaul@aisec.fraunhofer.de}
  }
  \and
  \IEEEauthorblockN{2\textsuperscript{nd} Alexander Küchler}
  \IEEEauthorblockA{\textit{Fraunhofer AISEC}\\
    Garching, Germany\\
    \href{mailto:alexander.kuechler@aisec.fraunhofer.de}{alexander.kuechler@aisec.fraunhofer.de}
  }
  \and
  \IEEEauthorblockN{3\textsuperscript{rd} Christian Banse}
  \IEEEauthorblockA{\textit{Fraunhofer AISEC}\\
    Garching, Germany\\
    \href{mailto:christian.banse@aisec.fraunhofer.de}{christian.banse@aisec.fraunhofer.de}
  }
}

\maketitle

\begin{abstract}
	The emergence of quantum computing raises the question of how to identify
	(security-relevant) programming errors during development. However, current
	static code analysis tools fail to model information specific to quantum
	computing.
	In this paper, we identify this information and propose to extend
	classical code analysis tools accordingly.
	Among such tools, we identify the \emph{Code Property Graph} to be very well suited for this task as it can be easily extended with quantum computing specific information.
	For our proof of concept, we implemented a tool which includes information
	from the quantum world in the graph and demonstrate its ability to analyze
	source code written in \emph{Qiskit} and \emph{OpenQASM}. Our tool brings
	together the information from the classical and quantum world, enabling
	analysis across both domains.
	By combining all relevant information into a single detailed analysis, this powerful tool can facilitate tackling future quantum source code analysis challenges. %
\end{abstract}

\begin{IEEEkeywords}
	static code analysis, software security, quantum code property graph, quantum
	source code analysis
\end{IEEEkeywords}

\section{Introduction}
In recent years, quantum computing started to gain relevance in research and
the first real-world implementations of quantum computers
have been proposed by industry \cite{IBMOsprey,arute2019google}.
Since quantum computing is a fairly new paradigm and
developers still need to adjust to the way quantum programs work, one can expect
an increasing number of implementations to be error-prone or even vulnerable.

This calls for the need to ensure the quality, safety and security of upcoming
quantum programs, e.g. mandated by the ``Talavera Manifesto''
\cite{piattini2020talavera}. Piattini et al. \cite{Piattini2020} formulate that
security and privacy by design and the quality of quantum software are key
aspects of Quantum Software Engineering.
While researchers
proposed numerous program analysis techniques for the security and safety
of classical computer programs for decades, there is a distinct lack of
methodologies to analyze quantum computing code.
Currently, quantum programs consist of a \classical{} and a \quantum{}.
This
requires an analysis across the boundaries of the two worlds, as proposed
by Valenica et al. \cite{valencia2021hybrid}.

However, due to the early stage of development, neither the bugs and
security implications nor the platform technologies of future quantum computing
programs can be completely foreseen. Therefore, a generalized approach to the
analysis of quantum programs, as well as their interface to the \allClassical{}
is needed. This approach must be independent of the actual quantum
programming language, since the popularity of languages might change in the
future, and needs to offer an extensible way to perform analysis tasks %
for problems which might not yet be known.

In this paper, we propose a static code analysis technique which can be applied
to both quantum and classical programs. We believe that an analysis
platform for quantum programs should allow
maximal flexibility with a minimal loss of information.
One such technique %
is the so-called \ac{cpg} \cite{ModelingAndDiYamagu2014}, a language-independent
graph model combining multiple graphs used in static code analysis.
We extend this concept to quantum computing programs by modeling the memory and
operations as well as dependencies between the qubits or quantum registers. We
also bridge the boundary between the \whole{} of the program.
In summary, our contributions are as follows:
\begin{itemize}
      \item We propose a static program analysis technique which spans the
            \whole{}. Our prototype 
            implementation\footnote{\href{https://github.com/Fraunhofer-AISEC/cpg/tree/quantum-cpg}{https://github.com/Fraunhofer-AISEC/cpg/tree/quantum-cpg}} 
            can analyze \qiskit{} and
            \qasm{}.
      \item We iterate different use-cases for bugs in quantum computing
            programs and present graph queries to identify them using our
            prototype. Lastly, we show the calculation of complexity metrics
            using our approach.
\end{itemize}

\section{Background}
In this section, we present the necessary background on code analysis and
quantum programming.

\subsection{Code Property Graphs}

A \ac{cpg} builds upon previous graph-based program analysis techniques which
use graph traversals to analyze the code. %
Yamaguchi et al. \cite{ModelingAndDiYamagu2014} combine multiple graphs such as the \ac{ast},
\ac{pdg} and \ac{cfg} into a single supergraph.
Since then, numerous frameworks have been proposed by security analysts, e.g.,
the open-source project \textit{cpg} \cite{weiss2022languageindependent}. We
chose this implementation since it is extensible and provides the ability to
analyze Python code which is prevalent in current quantum computing frameworks.
The translation of code is split into so-called \textit{language frontends}
which transfer the source code to the \ac{cpg}'s \ac{ast} representation and
\textit{passes} which enrich the \ac{cpg} with more information like \ac{dfg}
and \ac{eog} edges, among others. %

\subsection{Quantum Programming Languages}
\subsubsection{\qiskit{}}
\qiskit{} \cite{qiskit} is a popular open-source development framework for
quantum programs implemented in Python. It offers its users a
domain-specific language (DSL) to create quantum circuits. The circuit is
generated as a Python object, the qubits are represented as indices and the
gates are methods called on the circuit and receiving the qubits.
The API is designed to allow interactions between the quantum computing code and
the classical code (see Listing~\ref{lst:qiskit1}).

\begin{lstlisting}[numbers=right,frame=none,
  xrightmargin=15pt,
  caption={A simple code snippet in \qiskit{} which defines a quantum circuit
  where a Hadamard-gate (Line 2) is applied to qubit 0 and the result is 
  measured to classical
  bit 0. The circuit is executed 1000 times.},
  label=lst:qiskit1,
  style=mypython,
  language=Python,
  frame=none,
  escapechar=§,
]
circuit = QuantumCircuit(1, 1)
circuit.h(0)§\label{lst:qiskit1:h}§
circuit.measure([0], [0])§\label{lst:qiskit1:measure}§
compiled = transpile(circuit, simulator)
job = simulator.run(compiled, shots=1000)
result = job.result() # Fetch the results
counts = result.get_counts(compiled)
\end{lstlisting}

\subsubsection{\qasm{}}
\qasm{} \cite{qasm} is a
low-level language for quantum computers. The language features
many concepts of classical computing like branching, %
comparisons, function calls and classical types. \qasm{} extends this with
quantum computing features by adding specialized types, qubits, quantum
gates, measurements and reset instructions. \qasm{} does not handle the
execution of the program but only defines the code of the quantum service.
Executing QASM code can e.g. be scheduled in \qiskit{}.

\begin{lstlisting}[
  caption={The same circuit as in Listing
  \ref{lst:qiskit1} but defined with \qasm{}. %
},
  xrightmargin=15pt,
  label=code:qasm1,
  style=mypython,
  frame=none,
  numbers=right
]
qreg q[1];
creg c[1];
h q[0];
measure q[0] -> c[0];
\end{lstlisting}

\subsection{Architecture of Quantum Programs}

\begin{figure}[tb]
  \centering
  \resizebox{\columnwidth}{!}{
    \begin{tikzpicture}
      \node[part] (LOCAL) { \xcapitalisewords{\local{}} };
      \node[part, right = 1cm of LOCAL] (CLASSICAL) { \xcapitalisewords{\remote{}} };
      \node[part, right = 1cm of CLASSICAL] (QUANTUM) { \xcapitalisewords{\quantum{}} };

      \draw[very thick, dashed, red, rounded corners] ($(LOCAL.north west) + (-0.2cm, 0.55cm)$) rectangle ($(CLASSICAL.south east)+(0.2cm, -0.2cm)$);
      \node[red,above=0.05cm of LOCAL.north] { \xcapitalisewords{\allClassical{}} };

      \draw[very thick, dotted, black!60!green, dotted, rounded corners] ($(CLASSICAL.north west) + (-0.2cm, 0.7cm)$) rectangle ($(QUANTUM.south east)+(0.2cm, -0.35cm)$);
      \node[black!60!green,above=0.15cm of QUANTUM.north] { \xcapitalisewords{\quantumService{}} };

      \draw[thick, blue] ($(LOCAL.north west) + (-0.4cm, 1.5cm)$) rectangle ($(LOCAL.south east)+(0.4cm, -0.45cm)$);
      \node[blue,above=0.7cm of LOCAL.north] { \shortstack[c]{Classical\\Local Machine} };

      \draw[thick, blue] ($(CLASSICAL.north west) + (-0.4cm, 1.5cm)$) rectangle ($(CLASSICAL.south east)+(0.4cm, -0.45cm)$);
      \node[blue,above=0.7cm of CLASSICAL.north] { \shortstack[c]{Classical Hardware\\in the Datacenter} };

      \draw[thick, blue] ($(QUANTUM.north west) + (-0.4cm, 1.5cm)$) rectangle ($(QUANTUM.south east)+(0.4cm, -0.45cm)$);
      \node[blue,above=1.0cm of QUANTUM.north] { Quantum Computer };
    \end{tikzpicture}}
  \caption{Our understanding of different components of quantum programs (in black) and the hardware on which they run (blue boxes).}
  \label{fig:architecture}
\end{figure}
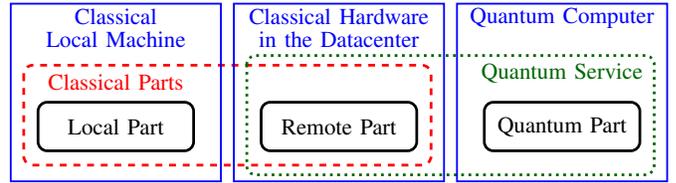

Quantum programs consist of multiple
parts which interact with each other and which run on different machines.
This impacts the design of quantum programs and lets us distinguish the
following components as shown in Figure \ref{fig:architecture}:
\begin{itemize}
  \item The \emph{\local{}} is run on the user's local machine and follows
        the classical computing paradigms.
  \item The \emph{\remote{}} runs %
        in the
        datacenter operating the quantum computer. It performs classical computing but
        interacts with the quantum parts through built-in APIs.%
  \item The \emph{\quantum{}} is the code which is actually executed on the
        quantum computer hardware.
  \item The \emph{\quantumService{}} embeds the remote \classical{} and the
        \quantum{} (i.e., it consists of the whole circuit including potential
        context switches).
  \item The \emph{\allClassical{}} of the program are all parts ran on
        classical hardware and include the local and remote part.
\end{itemize}

\section{Requirements}
\label{sec:requirements}
Based on the current state of development of the quantum computing ecosystem,
we identified the following requirements for
our analysis platform.

\vspace{1ex}
\noindent\textbf{R1: Adaptability to Future Changes.}~
As the technological stack of quantum computing is still in a relatively early
stage of development, the platform should provide enough flexibility to respond
to future changes, including new programming languages or software
architectures. Access to existing information should not suffer from such
changes.
This requires a certain level of abstraction from current implementations.

\vspace{1ex}
\noindent\textbf{R2: Considering Classical and Quantum Parts.}~ Currently,
quantum programs are developed as a mixture of classical and quantum parts where
both parts interact with each other through well-defined interfaces. The way in
which a quantum part is embedded in the classical part can provide interesting
insights to an analyst and thus has to be modeled. If similar information is
available in the classical and quantum parts of a program, it should be
accessible through a common interface to simplify the usage of the analysis 
platform.

\vspace{1ex}
\noindent\textbf{R3: Scalability.}~
Current implementations of quantum programs are still limited with respect to
the
number of qubits or gates and have a moderate complexity. However, with the
foreseeable advancements in the practicability of the
technologies, the quantum circuits will get more complex. This,
in turn, will require a highly scalable analysis platforms.

\section{Design}
\label{sec:design}
We propose to model a quantum program as an extension of a 
\ac{cpg}.
We call this new quantum-related part of the graph \ac{qcpg}. As the 
\classical{} of the \ac{cpg} remains unchanged, we can keep existing analysis 
techniques. We believe that a \ac{cpg} is a suitable representation as
it can abstract from the actual code with a minimal loss of 
information. Additionally, the representation allows to extend the model with 
more information without the need to change established interfaces to the 
\ac{cpg}. This makes the representation future-proof and allows responding to 
upcoming changes in the technological stack.

\subsection{Extending the Graph Model}
Our model extends the \ac{cpg}, which has been designed to analyze classical
software systems with semantic information related to quantum circuits. Similar
to prior work which aims at extending UML diagrams \cite{perez2021modelling,
  jimenez2021kdm}, we add concepts of \textit{circuit}s, \textit{qubit}s,
\textit{gate}s and \textit{measurement}s. However, in contrast to the proposed
extensions to UML profiles, we distinguish between different types of gates.
Furthermore, we add the concept of classical bits. The classical bits are used
to exchange data between the circuit and the classical parts of the
program.

We model each of our concepts as a separate type of node in the graph. The
nodes are connected to each other with edges which either represent an input to
a gate (including its order), or that a qubit is measured to a classical bit.
Overall, this results in a graph holding classical information and one
holding the information related to quantum computing code.

Figure \ref{fig:graph:example1} shows a part of the graph derived from
Listing \ref{lst:qiskit1}. The dashed line depicts the border
between classical programming (at the top) and the new quantum computing
semantics.

\subsection{Interactions between the Classical and Quantum Parts}

Most current algorithms are implemented in a hybrid way, i.e., they consist of 
\whole{}. The \local{} pre-processes data, interprets
the results of the circuit and parametrizes and executes
the circuit in a way to retrieve the desired results. This requires all parts
to interact with each other through dedicated means. With the concept of
classical bits, we already introduce the medium to exchange data between
both parts of the program. %
Executing the circuit and passing data to or retrieving the result from a
circuit's runs happens through a small set of well-known functions or 
operations. This makes it easy to identify such points where we add transitions 
from the
\ac{qcpg} to the classical \ac{cpg}.

\begin{figure}[t]
  \includegraphics[width=\columnwidth]{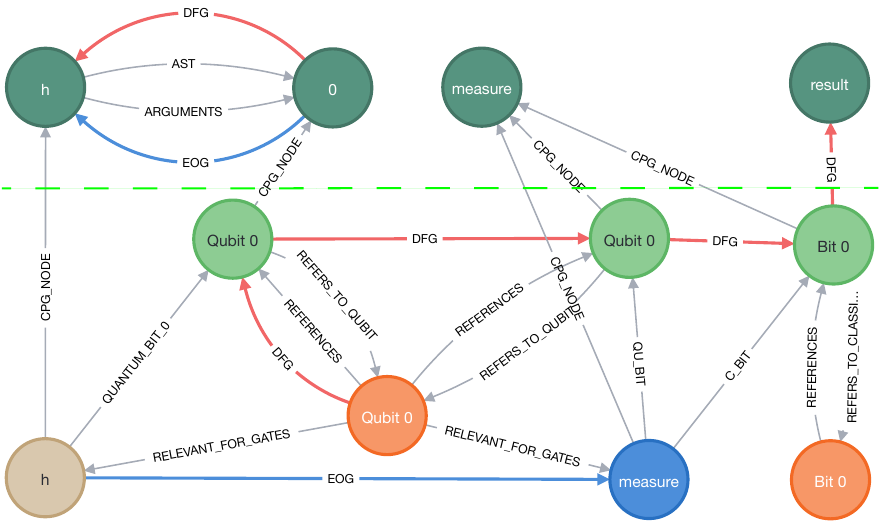}
  \caption{The (partial) CPG corresponding to Listing \ref{lst:qiskit1}. The
    circuit is generated with nodes above the dashed green line (the classical 
    \ac{cpg}) in \qiskit{'s} Python DSL (dark green nodes). We add the semantic 
    information and generate the \ac{qcpg} which additionally contains the nodes 
    below the dashed line. This graph consists of classical and quantum bits 
    (orange nodes) and the usages of the bits (light green nodes). It also 
    contains the Hadamard gate (beige node - Line \autoref{lst:qiskit1:h}), and 
    the measurement (blue node - Line \autoref{lst:qiskit1:measure}). The 
    evaluation order and dataflow edges span both the \whole{}
    of the program and let the parts interact with each other. I.e., the 
    \ac{dfg} edges show the dataflow to the call of \lstinline{result()} after 
    running the circuit.}
  \label{fig:graph:example1}
\end{figure}

Similar to the classic \ac{cpg}, we add \ac{dfg} and \ac{eog} edges to
our graph. This is shown by the red and blue edges in Figure
\ref{fig:graph:example1}. These edges flow across the
classical-quantum boundary.

Recently, the term ``dynamic circuits'' has been used to include controlflow
modifying code such as if-statements or loops into the languages used for quantum
programming. Essentially, these instructions measure a qubit, make a
branching decision and continue computing in the \quantum{} of the algorithm
based on the condition.
To account for such techniques, we add a branching node with the test condition 
and a branch statement to the \ac{qcpg}. %
This ensures that we do not have to split the circuit and
yet includes this controlflow in our graph. This emphasizes the relevance
of an analysis pipeline which considers both parts of the program.

Table~\ref{tab:nodes} summarizes all node types of the graph. As our
extensions only add information to the original \ac{cpg}, we can
query the graph for the code related to the \quantumService{}, while
completely reusing existing queries for the \classical{}. Additionally, we can 
track dataflow across
all parts of the code.

\begin{table}
  \caption{Node types in the Quantum Graph}
  \label{tab:nodes}
  \begin{tabular}{|l|p{5.5cm}|}
    \hline
    \textit{QuantumCircuit}      & Container for classical and quantum bits and one or more operations, which are executed remotely in the quantum computer or its quantum-classical-interface.                                                          \\\hline
    \textit{QuantumGate}         & A gate operation as part of the \textit{QuantumCircuit}, for example a Hadamard gate (\textit{QuantumGateH}).                                                              \\\hline
    \textit{QuantumBit}          & Declaration of a qubit inside a circuit.    \\\hline
    \textit{QuantumBitReference} & Reference to a \textit{QuantumBit}. \\\hline
    \textit{QuantumRegister}     & A register spanning multiple \textit{QuantumBit}s.     \\\hline
    \textit{QuantumMeasure}      & A single measurement from one \textit{QuantumBit} to a
    \textit{ClassicBit}. Multiple measurements are split up into separate nodes. \\ \specialrule{.2em}{0pt}{0pt} %
    \textit{ClassicBit}          & Single piece of memory (1 bit) in the quantum-classic interface of a quantum computer.                                      \\\hline
    \textit{ClassicBitReference} & Reference to a \textit{ClassicBit}.                                                                          \\\hline
    \textit{ClassicRegister}     & A register spanning multiple \textit{ClassicBit}s.     \\\hline
    \textit{ClassicIf}           & A control structure inside the quantum-classic interface, that operates on a measured \textit{ClassicBit} and allows the dynamic execution of \textit{QuantumGate}s based on the comparison.                    \\\hline
  \end{tabular}
\end{table}

\subsection{\ac{eog}}
\label{sec:EOGpass}
Once the quantum nodes have been created, we can (language-independently) add
\ac{eog} edges. We achieve this by traversing the \ac{cpg} for \emph{circuit}
declarations and then follow the operations on the circuit in order, adding 
\ac{eog} edges along the way. Control
structures (\emph{ClassicIf}, ...) are considered test-expression first, then
the body. There is an \ac{eog} flow from the test-expression to the body (the 
test is \emph{true}) and a flow to the next instruction (the test is 
\emph{false}).
Overall, the \ac{eog} is constructed similarly established as for classical programs.

\subsection{\ac{dfg}}
\label{sec:DFGpass}
The \ac{dfg} depends on the \ac{eog} and is more interesting to construct. We
keep the \ac{dfg} out of the \emph{QuantumGate}s, as our operations make use of
\emph{QuantumBitReference}s instead of the \emph{QuantumBit}s directly. This
enables us to track individual dataflows.
Specifically, we first loop through all operations and connect their
\emph{QuantumBitReference}s internally (i.e. we add a \ac{dfg} from the control
bit to the controlled bit for the \emph{CX} gate). We then identify the first
operation acting on each qubit and traverse the (potentially multiple) \ac{eog}
edges from there, adding \ac{dfg} edges along the way, thus connecting the operations
with each other.
Finally, we reconnect the classic \ac{cpg} to the \ac{qcpg} by connecting the
\emph{measure} instruction with the \qiskit{} \lstinline{result} instruction
(and subsequent bit sensitive access to it), thus providing dataflow edges from
the quantum \emph{ClassicBit} node back to the classic \ac{cpg} node.

\section{Implementation}

We implemented this \ac{qcpg} using the open source \cpg{} library
\cite{weiss2022languageindependent}.
We summarize implementation details of the \textit{language
 frontends} and \textit{passes} which %
parse \qiskit{} and \qasm{} code and %
transfer the information into %
the graph model. 

\subsection{Parsing QC Code}
As a first step, we extended the \cpg{} with the ability to handle \qiskit{} and
\qasm{} code.
\qiskit{} easily translates to a classic \ac{cpg}, as it is pure Python code.
E.g., the instruction \lstinline{qc.cx(0, 1)} translates to 
a \emph{CallExpression} to the function \lstinline{cx} receiving the integer
parameters \lstinline{0} and \lstinline{1}.

For \qasm{}, we implemented an OpenQASM v3 parser since we are not aware of any 
functional off-the-shelf implementation which provides us with an \ac{ast}. Our
parser translates the source code to the \cpg{}'s version of the \ac{ast}.
Mapping the low-level instructions of \qasm{} to the high-level \ac{cpg}
representation is not straight-forward. Most interestingly, we
map gate instructions to function calls. Arithmetic instructions and controlflow
modifying instructions such as \lstinline{if}-statements directly map to the
classical \ac{cpg}. For \lstinline{reset} and \lstinline{measure} instructions,
we use \emph{CallExpression}s just as in \qiskit{}.

\subsection{Quantum Code Property Graph}
The challenge is to add the extra semantic information (Section \ref{sec:design})
to the graph. We implemented a pass adding the quantum nodes
described in Table \ref{tab:nodes} to the classical \ac{cpg} and thus generating
the \ac{qcpg}. The pass further connects the nodes of the \whole{} of
the resulting graph.

For \qiskit{}, we scan the \ac{cpg} for variables which hold
the return value of the method \lstinline{QuantumCircuit}. This lets us create
the \emph{QuantumCircuit}, \emph{QuantumBit} and \emph{ClassicBit} nodes for
our \ac{qcpg}. We then traverse the \ac{cpg} to collect all \emph{CallExpression}s
on the variable to find all operations and create the respective nodes (e.g.,
\emph{QuantumGate}, \emph{QuantumMeasure}, ...).

For the instruction \lstinline{qc.h(0)} in Listing 
\ref{lst:qiskit1}, this results in
the following nodes in Figure \ref{fig:graph:example1}:
\begin{itemize}
  \item a \emph{QuantumGateH} (beige node)
  \item a \emph{QuantumBitReference} (light green node) connected to qubit 0 and 
    the Hadamard gate
  \item a \emph{QuantumBit} (orange node) connected to its references
\end{itemize}

This results in quantum nodes which are now enriched with EOG and DFG edges
as described in Sections \ref{sec:EOGpass} and \ref{sec:DFGpass}.

\section{Usage Scenarios}
\label{sec:usage}

In this section, we demonstrate the effectiveness of our approach and
illustrate the capabilities of our analysis platform. %
Due to the
lack of large-scale datasets that contain security or programming errors, we
tried to come up with potential programming errors and code smells and created
example source code with these flaws in Qiskit and
OpenQASM. Partially, these flaws are inspired by flaws usually found in 
classical computing and are adapted to their quantum counterparts. Other
flaws originate from the distinct interfaces between the \remote{} and \quantum{}%
, as well as
interactions with the final result provided by a \quantumService{} to the
local part of the code. %
Lastly, we also provide exemplary queries
to calculate different complexity metrics on our dataset.

\begin{lstlisting}[label=lst:inputoutput,style=mypython,language=Python,numbers=right,frame=none,xrightmargin=20pt,
  caption={Listing for \texttt{complex.py}. In Line 
  \autoref{lst:inputoutput:localvars}, two variables representing individual 
  bits are defined.
  These are then used to initialize the state vector of different qubits
  (Line \autoref{lst:inputoutput:initStart}-\autoref{lst:inputoutput:initEnd}). 
  Lines \autoref{lst:inputoutput:opsStart}-\autoref{lst:inputoutput:opsEnd} 
  represent several quantum operations, such as gate
  operations and measurements. See Figure~\ref{fig:circuit-ineffective}
  for a visualization of these operations. Lines 
  \autoref{lst:inputoutput:transpile}-\autoref{lst:inputoutput:run} contain code 
  that executes the quantum code (either on
  hardware or a simulator) and Lines 
  \autoref{lst:inputoutput:result}-\autoref{lst:inputoutput:counts} fetch the 
  result from the
  \quantumService{} to a local variable. The local result is then split back 
  into individual bits in Lines 
  \autoref{lst:inputoutput:accessStart}-\autoref{lst:inputoutput:accessEnd}.
  Finally, a (local) code execution decision is made based on one measured
  bit in Lines \autoref{lst:inputoutput:if}-\autoref{lst:inputoutput:complex}.},
  escapechar=§]
a = 0; b = 1§\label{lst:inputoutput:localvars}§
# Create a quantum circuit with 4 qubits
circuit = QuantumCircuit(4, 4)
# Initialize qubits with local part values
circuit.initialize(str(a), 0)§\label{lst:inputoutput:initStart}§
circuit.initialize(str(a), 1)
circuit.initialize(str(b), 2)§\label{lst:inputoutput:initQubit2}§ 
circuit.initialize(str(b), 3)§\label{lst:inputoutput:initEnd}§
# Different quantum operations
circuit.h(0); circuit.h(3)§\label{lst:inputoutput:opsStart}§
circuit.cx(1, 0)
circuit.measure([2], [2])
circuit.x(1).c_if(2, 0)
circuit.measure([1,3], [1,3])§\label{lst:inputoutput:opsEnd}§
# Run job on target (hardware or simulator)
cc = transpile(circuit, target)§\label{lst:inputoutput:transpile}§
job = target.run(cc, shots=1000)§\label{lst:inputoutput:run}§
# Grab results from the job
result = job.result()§\label{lst:inputoutput:result}§
counts = result.get_counts(cc)§\label{lst:inputoutput:counts}§
# Evaluate results back in local part
for bitstring in counts:
  c0 = int(bitstring[-0-1])§\label{lst:inputoutput:accessStart}§
  c1 = int(bitstring[-1-1])
  c2 = int(bitstring[-2-1])
  c3 = int(bitstring[-3-1])§\label{lst:inputoutput:accessEnd}§
  # Execute code based on a decision
  # of a measured bit for every result
  if c2 == 1:§\label{lst:inputoutput:if}§
    do_something_complex()§\label{lst:inputoutput:complex}§
\end{lstlisting}

Listing~\ref{lst:inputoutput} shows an example Qiskit source code which
demonstrates the interaction between the \quantumService{} and the \local{} of the program.

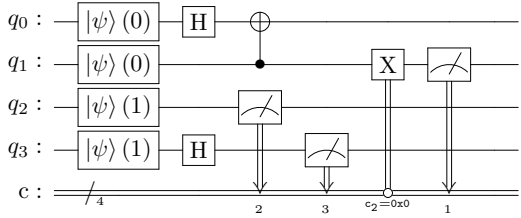
\begin{figure}[h!]
  \centering
  \scalebox{0.9}{
  \Qcircuit @C=1.0em @R=0.2em @!R { \\
  \nghost{{q}_{0} :  } & \lstick{{q}_{0} :  } & \gate{|\psi\rangle\,(\mathrm{0})} & \gate{\mathrm{H}} & \targ & \qw & \qw & \qw & \qw & \qw\\
  \nghost{{q}_{1} :  } & \lstick{{q}_{1} :  } & \gate{|\psi\rangle\,(\mathrm{0})} & \qw & \ctrl{-1} & \qw & \gate{\mathrm{X}} & \meter & \qw & \qw\\
  \nghost{{q}_{2} :  } & \lstick{{q}_{2} :  } & \gate{|\psi\rangle\,(\mathrm{1})} & \qw & \meter & \qw & \qw & \qw & \qw & \qw\\
  \nghost{{q}_{3} :  } & \lstick{{q}_{3} :  } & \gate{|\psi\rangle\,(\mathrm{1})} & \gate{\mathrm{H}} & \qw & \meter & \qw & \qw & \qw & \qw\\
  \nghost{\mathrm{{c} :  }} & \lstick{\mathrm{{c} :  }} & \lstick{/_{_{4}}} \cw & \cw & \dstick{_{_{\hspace{0.0em}2}}} \cw \ar @{<=} [-2,0] & \dstick{_{_{\hspace{0.0em}3}}} \cw \ar @{<=} [-1,0] & \controlo \cw^(0.0){^{\mathtt{c_2=0x0}}} \cwx[-3] & \dstick{_{_{\hspace{0.0em}1}}} \cw \ar @{<=} [-3,0] & \cw & \cw\\
  \\ }}

  \caption{The quantum circuit contained in Listing~\ref{lst:inputoutput}.}
  \label{fig:circuit-ineffective}
\end{figure}

\subsection{Queries to find programming errors}

In order to find patterns that correspond to programming errors or code smells,
we analyzed the source code in Listing~\ref{lst:inputoutput} using our prototype
and stored the resulting graph in a Neo4J graph database. In the following, we
discuss several graph queries to find the following patterns.

\subsubsection{\textbf{SUPERFLUOUS\_OPERATION}: Quantum gate does not affect
  measurement}

In this scenario we are interested in quantum operations, e.g., gates, that
affect qubits, but those qubit are not measured or do not affect other qubits
which are measured. This makes the original gate operation superfluous and can
be regarded as a programming error.

\begin{lstlisting}[breaklines=true,language=sql,label=lst:op-not-measured,caption=Cypher 
query to find superfluous operations by identifying 
\textit{QuantumBitReference}s which do not have a dataflow to a 
\textit{QuantumMeasure}.,
language=cypher]
MATCH p=(:QuantumGate)-->(r:QuantumBitReference)
WHERE NOT EXISTS{(r)-[:DFG*]->(:QuantumBitReference)
<-[:QU_BIT]-(:QuantumMeasure)} RETURN p
\end{lstlisting}

Listing~\ref{lst:op-not-measured} shows a possible Cypher (a query language for
Neo4j graph databases) query to find such
gate operations. Specifically, we look for \textit{QuantumGate} nodes that do
not have a dataflow from any of their qubit arguments to a
\textit{QuantumMeasure} node. Given the circuit in
Figure~\ref{fig:circuit-ineffective}, the query returns that gates
\textit{H} and \textit{CX} operating on $q_0$ and $q_1$ are ineffective on the
result. While the \textit{CX} gate does have a dataflow towards $q_0$ (its
target qubit), the state vector of $q_1$ (its control qubit) is not changed and
is measured in its initial state.
Most likely the developer of this circuit made a mistake
in switching control and target qubit of the \textit{CX} operation or mistakenly
measured $q_1$ instead of $q_0$.

\subsubsection{\textbf{CONSTANT\_CLASSIC\_BIT}: A bit is measured but has not been transformed}
Besides not measuring $q_0$, the circuit in Figure \ref{fig:circuit-ineffective}
contains another potential bug because it measures $q_2$ which has
never been changed. This leads to an almost constant value for this bit which
does not carry much information but only increases the complexity of the
circuit. The query in Listing \ref{lst:const-measure} identifies such patterns.

\begin{lstlisting}[breaklines=true,language=sql,label=lst:const-measure,caption=Cypher 
query to find nearly constant measured qubits -- characterized by a direct 
dataflow from a \textit{QuantumBitReference} to a \textit{ClassicBitReference} 
without any other dataflow edges in between.,language=cypher]
MATCH p=(:QuantumBit)-[:DFG]->(:QuantumBitReference)
-[:DFG]->(:ClassicBitReference) RETURN p
\end{lstlisting}

\subsubsection{\textbf{CONSTANT\_CONDITION}: The condition of a classic-if
  always evaluates to 'true' or 'false'} This scenario is an extension of the
\textit{CONSTANT\_CLASSIC\_BIT} use-case. Recently, quantum computing
frameworks added the ability to include if statements in the quantum code.
This statement measures a qubit, compares it to another value (in the
classic part of the quantum service) and executes different branches based on
this result (in the quantum part). However, if the qubit to measure has not
been transformed by any operations (e.g., gates), it results in a nearly
constant evaluation. This indicates either missing transformations in the
circuit or, if the respective qubit is explicitly set as an input variable,
that the circuit should be split into distinct ones which are executed
depending on the classic variable which is used to initialize the qubit. Once
we execute the query in Listing~\ref{lst:const-condition}, we can determine that
the
\textit{ClassicIf} only depends on $c_2$ which is measured on $q_2$, but has
no operations affecting it (see previous scenario).

\begin{lstlisting}[breaklines=true,language=sql,label=lst:const-condition,caption={Cypher 
query to find nearly constant conditions by identifying \textit{ClassicIf} nodes which depend on bits that are not modified by gates.},language=cypher]
MATCH p=(a:ClassicIf)-[:CONDITION]->()-[:LHS]->
(r:ClassicBitReference) WHERE NOT EXISTS {
(r)<-[:DFG*3..]-(:QuantumBit)}
AND NOT EXISTS {(r)<-[:DFG*]-(:QuantumBit)
-[:RELEVANT_FOR_GATES]-(:QuantumGate)} RETURN p
\end{lstlisting}

\subsubsection{\textbf{RESULT\_BIT\_NOT\_USED}: Single bit measured but not used locally}

This scenario builds upon the observation that the result returned by the
\qiskit{} framework has to be decomposed and interpreted by the user. We assume
that this happens by accessing individual classic bits $c_X$ which are contained
in the key of the result dictionary. In this case, if a user does not access one
of the $c_X$ which are measured in the circuit, there might be a misfit between
the circuit and its usage in the program. We can simulate this by commenting out
one of the Lines
\autoref{lst:inputoutput:accessStart}-\autoref{lst:inputoutput:accessEnd} in
Listing \ref{lst:inputoutput}. This means that the measure
and the preceding operations are irrelevant to the further outcome of the
program and it might be worth optimizing the circuit. Listing
\ref{lst:storage-calls} shows a cypher query which detects such issues. This
use-case illustrates the need for an analysis across both parts of the
program.

\begin{lstlisting}[breaklines=true,language=sql,label=lst:storage-calls,caption=Cypher 
query to find unused result bits; they are unused if they the respective index is not accessed in the final bitstring result array.,language=cypher]
MATCH p=(r:ClassicBitReference)<-[:C_BIT]-
(:QuantumMeasure) WHERE NOT EXISTS {(r)-[:DFG]
->(:ArraySubscriptionExpression)} RETURN p
\end{lstlisting}

\subsubsection{\textbf{CONSTANT\_RESULT\_BIT}: Result bit is constant}

Building upon the previous dataflows, we can also detect if result bits are
used, but contain classic bits that were not affected by operations in the
quantum circuit. The query in Listing~\ref{lst:flow} traces
the DFG from the local function \lstinline{do_something_complex} (Line
\autoref{lst:inputoutput:complex} in
Listing~\ref{lst:inputoutput}) back to its origin in the \quantum{}. We can
see that the execution is based on the contents of $c_2$, which is measured on
$q_2$, but never changed by any gate operation. Its value is therefore still its
original state $1$, which was assigned by the local variable $b$ in Line
\autoref{lst:inputoutput:initQubit2}.

\begin{lstlisting}[breaklines=true,language=sql,label=lst:flow,caption={Query to trace back the execution of a \textit{CallExpression} to an \textit{IfStatement}, and then finding the source of the dataflow to that statement.},language=cypher]
MATCH p=(c:CallExpression)<-[:EOG*]-(:IfStatement)
-[:DFG*]-(:ArraySubscriptionExpression)
<-[:DFG]-(:ClassicBitReference)
<-[:DFG*]-(:QuantumNode)
WHERE c.name = "do_something_complex" RETURN p
\end{lstlisting}

\subsection{Complexity Metrics}

Cruz et al. \cite{cruz2021towards} presented several metrics to compare the
complexity of quantum circuits. We model these metrics as
queries (see Listing~\ref{lst:metrics}) to fetch complexity information about a
circuit automatically. This even works independently of the actual
quantum programming language used.

\begin{lstlisting}[breaklines=true,language=sql,label=lst:metrics,caption=Query to calculate a subset of the metrics proposed in 
\cite{cruz2021towards}.,language=cypher]
MATCH (q:QuantumBit) RETURN COUNT(q) AS value, 
  "Width" AS key UNION
MATCH (p:QuantumGate)<-[:RELEVANT_FOR_GATES]->(b:QuantumBit) RETURN COUNT(p) AS value, 
  "Depth" AS key ORDER BY value DESC LIMIT 1 UNION
MATCH (p:QuantumGate) RETURN COUNT(p) AS value,
  "NoGates" AS key UNION
MATCH (p:QuantumGateX) RETURN COUNT(p) AS value, 
  "NoP-X" AS key UNION
MATCH (p:QuantumGateY) RETURN COUNT(p) AS value, 
  "NoP-Y" AS key UNION
MATCH (p:QuantumGateZ) RETURN COUNT(p) AS value, 
  "NoP-Z" AS key UNION
MATCH (p:QuantumPauliGate) RETURN COUNT(p) AS value, 
  "TNo-P" AS key UNION
MATCH (p:QuantumGateH) RETURN COUNT(p) AS value, 
  "NoH" AS key UNION
MATCH (q:QuantumBit)-[:DFG]->(:QuantumBitReference)
  <-[:QUANTUM_BIT_0]-(:QuantumGateH) 
  WITH COUNT(q) AS countH
MATCH (b:QuantumBit) WITH COUNT(b) AS total, countH
  RETURN countH*1.0/total AS value, "%
\end{lstlisting}

\section{Related Work}

Prior work aims to model quantum circuits or quantum
programs. Jim\'{e}nez et al. \cite{jimenez2020reverse} generate a Knowledge 
Discovery Metamodel from Q\# code and transfer it to a quantum computing-specific
extension of the UML representation \cite{jimenez2021kdm}.
Similarly, other research \cite{perez2021modelling, delgado2020towards}
proposes to model quantum programs in UML graphs. Burgholzer et al.
\cite{burgholzer2021efficient} aim to extract blocks with similar behavior from
quantum circuit to ease further analysis.
\qiskit{} itself provides a tool to
visualize quantum circuits but neither allows for automated
analysis on the generated images nor encodes the connection to the \local{}.
All these models are unsuitable for an automated analysis.

Other research models quantum programs as finite state machines
\cite{moore2000quantum,say2014quantum,tian2019experimental}, via abstract
interpretation \cite{yu2021quantum} or aim to apply SAT solving to some classes
of quantum circuits \cite{berent2022towards}. These works require a specialized
encoding of the quantum circuits which, in many cases is limited in terms of
scalability \cite{shaukat2020modeling} and generalizability
\cite{berent2022towards}. Gheorghiu et al. \cite{gheorghiu2019verification}
summarize techniques which aim at verifying quantum programs.
However, researchers found that many bugs in quantum programs originate
from the boundary between the \whole{} and from the quantum computing frameworks
\cite{zhao2021identifying,paltenghi2022bugs}.

Cruz-Lemus et al.
\cite{cruz2021towards} propose a set of metrics for quantum circuits. We showed
that our tool can automatically measure these metrics for a quantum program.
Another branch of research aims to measure the maximal complexity of a circuit
which can be executed on a given quantum computer
\cite{salm2020about,bishop2017quantum,amico2023defining,cross2019validating,sete2016a}.
Our tool could automatically calculate these metrics which allows an analyst to compare it with
the achievable values of the hardware to identify potential problems with the
decoherence.
Yet other researchers aim at describing the errors of individual
gates \cite{nielsen2002a,willsch2017gate} on a physical level.
Zhao \cite{zhao2021quantum} provides an overview over these metrics, bug types, and
analyses of quantum programs.
While these metrics can help to identify potentially error-prone quantum
circuits, prior work does not allow users to automatically measure these metrics
and furthermore they do not account for hybrid concepts.

\vspace{1ex}
\noindent\textbf{Code Property Graphs.}~
A wide range of works focuses on applying \ac{cpg}s %
to different programming
languages \cite{ModelingAndDiYamagu2014,weiss2022languageindependent,joern},
low-level code representations or binaries
\cite{kuechler2022representing,schuette2019lios}, Java bytecode
\cite{keirsgieter2020graft,plume} or even cloud applications
\cite{banse2021cloud}. Furthermore, a wide range of research uses these tools to
analyze the security of classical programs e.g. by graph queries
\cite{alKassar2022testability}, by applying machine learning techniques
\cite{xiaomeng2018cpgva,yamaguchi2015automatic} or by extracting privacy
properties from the graphs \cite{kunz2023privacy}.
However, none of these works can be used to model quantum programs as they lack
support for the frameworks and do
not model any specifics of these programs. We extend the concept by adding more
information which is not available in classical programs and furthermore connect
both parts.

\begin{table}[t]
  \centering
  \caption{Comparison of existing tools with our approach}
  \label{tab:comparison}
  \resizebox{\columnwidth}{!}{
    \begin{tabular}{|l|c|c|c|c|}\hline
                            & \textbf{\qiskit{}} & \textbf{UML models} & \textbf{\ac{cpg}} & \textbf{Our approach} \\\hline
      \specialrule{.075em}{0pt}{0pt}
      Quantum parts         & \yes               & \yes                & \no               & \yes                  \\\hline
      Classical parts       & \no                & \yes                & \yes              & \yes                  \\\hline
      \specialrule{.075em}{0pt}{0pt}
      Superfluous Operation & \yes               & \no                 & \no               & \yes                  \\\hline
      Constant Classic Bit  & \no                & \no                 & \no               & \yes                  \\\hline
      Constant Condition    & \no                & \no                 & \no               & \yes                  \\\hline
      Result Bit Not Used   & \no                & \no                 & \yes        			 & \yes                  \\\hline
      Constant Result Bit   & \no                & \no                 & \no               & \yes                  \\\hline
      \specialrule{.075em}{0pt}{0pt}
      Complexity metrics    & \no                & \no                 & \no               & \yes                  \\\hline
    \end{tabular}}
\end{table}

\section{Discussion and Future Work}
We revisit the requirements from Section \ref{sec:requirements}, discuss
limitations of our tool and summarize future research directions.

\vspace{1ex}
\noindent\textbf{Requirements.}~
The requirements \textbf{R1} and \textbf{R3} are inherently met by using a
\ac{cpg} as analysis platform as they scale well even
for large code bases (\textbf{R3}) and provide a language-independent code
representation which can be easily extended (\textbf{R1}). Our solution
combines the existing \ac{cpg} with new quantum nodes in the
\ac{qcpg}. It keeps the interface consistent across all parts, thus enabling a
cross-domain analysis (\textbf{R2}).

\vspace{1ex}
\noindent\textbf{Limitations.}~
The usage scenarios discussed in Section \ref{sec:usage} describe an initial set
of problems. However, we believe that the scientific community will identify
more errors in quantum programs once the technology is more established. These
problems can be modeled as a query to our graph. At the
same time, the identified scenarios may not fit all usages of quantum
algorithms.
Due to the lack of an extensive set of test programs, our experiments
are limited to only few examples.

\vspace{1ex}
\noindent\textbf{Future work.}~
Future work can support more quantum programming languages, improve
the level of detail for dataflows and introduce the concept of parallel
execution into the \ac{eog}. Once novel error types are identified, new queries
should be developed to identify them. Furthermore, including hardware-dependent
information into the \ac{qcpg} is left to future work.

\section{Conclusion}
We propose a method to analyze quantum and classical computing code in a
unified representation.
By adding quantum specific nodes and edges, we enable analysis of the
entire program without hindering existing analyses of the \local{} of the
program.
We further propose an initial set of analyses detecting
potential programming errors, following dataflows or calculating complexity
metrics of quantum programs.

\balance
\bibliographystyle{IEEEtran}%
\bibliography{refs}
\end{document}